**Insights into the Structure and Ion Transport of Pectin-[BMIM][$PF_6$] Electrolytes**

*Sipra Mohapatra*[1], *Sougata Halder*[1], *Sachin R. Chaudhary*[2], *Roland R. Netz*[3], *and Santosh Mogurampelly*[1,3,*]

[1]Polymer Electrolytes and Materials Group (PEMG), Department of Physics, Indian Institute of Technology Jodhpur, Karwar, Rajasthan 342037, India.

[2]Department of Spice and Flavour Science, CSIR-Central Food Technological Research Institute, Mysore, Karnataka 570020, India

[3]Fachbereich Physik, Freie Universität Berlin, 14195 Berlin, Germany.

*Corresponding Author: santosh@iitj.ac.in

## ABSTRACT

We investigate the effect of pectin on the structure and ion transport properties of the room-temperature ionic liquid electrolyte 1-n-butyl-3-methylimidazolium hexafluorophosphate ([BMIM][$PF_6$]) using molecular dynamics simulations. We find that pectin induces intriguing structural changes in the electrolyte that disrupt large ionic aggregates and promote the formation of smaller ionic clusters, which is a promising finding for ionic conductivity. Due to pectin in [BMIM][$PF_6$] electrolytes, the diffusion coefficient of cations and anions is observed to decrease by a factor of four for a loading of 25 wt. % of pectin in [BMIM][$PF_6$] electrolyte. A strong correlation between the ionic diffusivities ($D$) and ion-pair relaxation timescales ($\tau_c$) is observed such that $D \sim \tau_c^{-0.75}$ for cations and $D \sim \tau_c^{-0.82}$ for anions. The relaxation timescale exponents indicate that the ion transport mechanisms in pectin-[BMIM][$PF_6$] electrolytes are slightly distinct from those found in neat [BMIM][$PF_6$] electrolytes ($D \sim \tau_c^{-1}$). Since pectin marginally affects ionic diffusivities at the gain of smaller ionic aggregates and viscosity, our results suggest that pectin-IL electrolytes offer improved properties for battery applications, including ionic conductivity, mechanical stability, and biodegradability.

## INTRODUCTION

There is an increased demand for advanced solid polymer electrolytes (SPEs) for battery applications possessing high ionic conductivity at room temperature, better mechanical stability, and environmental compatibility.[1–5] As a portable energy storage device, lithium-ion batteries are becoming more dominant due to their unique characteristics, such as lightweight, high mechanical, and electrochemical stability.[6–11] Electrolytes are found in liquid, solid, and gel forms, serving as a medium for ion transport between the electrodes, and are a prime component of batteries.[12] Commercially available lithium-ion batteries commonly use carbonate-based organic solvents as liquid electrolytes because of their high dielectric constant (~90), melting point, and ionic conductivity (~$10^{-4}$ S/cm) at room temperature.[13] However, the liquid electrolytes are highly flammable, nonbiodegradable, and highly volatile, leading to battery failure due to charge leakages, dendritic growth, and short-circuiting.[14–16] To circumvent the drawbacks mentioned above, researchers focus on designing and synthesizing polymer-based electrolytes that are highly conductive, non-volatile, and mechanically stable.[11,17]

Solid polymer electrolytes (SPEs) are receiving more attention due to fulfilling the goals mentioned above and their electrochemical stability and cost-effectiveness.[16,18,19] SPEs generally suffer from a series of shortcomings that have delayed their application in commercial devices. These include several factors, such as crystallizing effects and significantly lower ionic conductivity (usually lower than $10^{-5}$ S/cm at room temperature) than liquid electrolytes.[20,21] SPEs are also prone to be bent, stretched out, and deformed when used in flexible batteries, leading to breakage and micro-cracks resulting in short circuits. Further, the SPEs typically suffer from low transference numbers, high cation-anion correlations, low ionic conductivity, and nonbiocompatibility.[4,22,23]

Recently, we simulated pectin-EC-LiTFSI electrolytes and investigated their structural and ion transport properties and reported that the pectin chains reduce the coordination of lithium ions with TFSI ions, leading to smaller ionic aggregates.[24] The diffusivity of lithium and $TFSI^-$ (bis(trifluoromethanesulfonyl)imide) ions was found to correlate inversely with the ion-pair relaxation timescales as $D \sim \tau_c^{-3.1}$, and $D \sim \tau_c^{-0.95}$, respectively, highlighting distinct transport mechanisms for the ionic species. That study suggests that pectin-based electrolytes could be a promising alternative to traditionally used synthetic solid polymer electrolytes.

The traditionally employed liquid EC-salt electrolytes in battery applications are typically organic solvents that contain dissolved salts, usually lithium salts.[25,26] They possess excellent ionic conductivity and can facilitate faster movement of ions between the battery electrodes. However, they also have some limitations, such as a relatively narrow electrochemical stability window, which can lead to unstable solid-electrolyte interphase (SEI) near the surface of the electrodes. The SEI layer can decrease the efficiency of the battery and reduce its capacity over time. In contrast to the EC-salt type of liquid electrolytes, ionic liquids (ILs) themselves act as liquid electrolytes consisting of only ionic species. They have unique properties, including non-volatility, non-flammability, and good thermal and electrochemical stability.[27–29] They can have a wider electrochemical stability window compared to EC electrolytes, leading to improved battery performance and stability.[29–31] Therefore, exploring pectin-IL electrolytes could be a promising avenue for developing new and improved electrolytes with optimized properties for energy storage applications.[32,33]

In this work, we hypothesize that loading pectin in an electrolyte composed of large ionic species as found in RTILs instead of the flammable liquid electrolytes (for example, EC) reduces the interactions between the cations and monomeric pectin groups and, thereby, addresses a few critical issues such as low transference numbers, the correlated motion of ions, and inefficient ionic conductivity.[34–37] We selected the [BMIM][$PF_6$] ionic liquid as our primary electrolyte due to its excellent properties, including high ionic conductivity, nonvolatility, and large molecular size for use in the pectin-containing system.[27,38] We hypothesize that [BMIM][$PF_6$] ionic liquid promotes high cation transference numbers in the electrolyte compared to pectin-EC-LiTFSI electrolytes due to the reduced correlations with monomeric pectin groups. We conducted simulations to investigate the ion transport and structural properties of different pectin-loaded systems (2, 5, 10, 20, and 25 wt. %) and compared them to neat [BMIM][$PF_6$] electrolytes (0 wt. %). The simulation details (including the forcefield, interaction potential, system setup, and equilibration protocol), structure, ion diffusion, ion-pair correlations, and conclusions are presented in the following sections.

We find that incorporating 25 wt. % of pectin in the electrolyte only leads to a moderate reduction in ion diffusivity, approximately by a factor of 4. This reduction is not substantial when compared to the pectin-EC-LiTFSI electrolyte systems. Our analysis reveal that the diffusivity of ionic species in pectin-[BMIM][$PF_6$] electrolytes strongly correlate with the ion-pair relaxation timescales such that $D{\sim}\tau_C^{-0.75}$ for cations and $D \sim \tau_C^{-0.82}$ for anions. Hence, the structural relaxations occurring due to the association and dissociation of the ion-pairs are the primary transport mechanisms observed in pectin-[BMIM][$PF_6$] along with additional effects likely coming from the pectin. Despite the slightly diminished ion diffusivity, the overall benefits of mechanical properties and biodegradability make it a favorable option for practical applications. Based on the results, we speculate that biopolymers can replace synthetic polymers to make the rechargeable battery electrolytes more biocompatible, highly stable, and efficient.

## SIMULATION DETAILS

### Interaction potential

Atomistic molecular dynamics simulations were performed on composite electrolytes having pectin chains solvated in pure [BMIM][$PF_6$] using the GROMACS 2020.2 package.[39] The interaction potential used for the simulations is,

$$U(r) = \frac{1}{2}k_r(r - r_0)^2 + \frac{1}{2}k_\theta(\theta - \theta_0)^2 + \frac{1}{2}\sum_{n=1}^{4} C_n[1 + (-1)^{n+1}\cos(n\phi)] + \sum 4\epsilon\left[\left(\frac{\sigma}{r_{ij}}\right)^{12} - \left(\frac{\sigma}{r_{ij}}\right)^{6}\right] + \sum \frac{q_1 q_2}{4\pi\epsilon_0 r_{ij}}$$

containing bonded and nonbonded terms, i.e., U(**r**) = $U^{bonded}$(**r**) + $U^{nonbonded}$(**r**). The bonded potential consists of the interaction arising from the bonds, angles, dihedrals, and improper torsions in the pectin-loaded system. Explicitly, we used a harmonic potential of the form $\frac{1}{2}k(r - r_0)^2$ for bonds, $\frac{1}{2}k(\theta - \theta_0)^2$ for angles, and $\frac{1}{2}\sum_{n=1}^{4} C_n[1 + (-1)^{n+1}\cos(n\phi)]$ for dihedrals. The nonbonded term consists of the standard LJ and Coulombic interaction potential. A scaling factor of 0.5 was used for nonbonded 1-4 interactions within a molecule.

For the IL, the OPLS-AA force field parameters were used for the bonded and the nonbonded interaction terms.[40] We further used the refined parameters for ILs developed by Bhargava et al.[41] along with the scaling of charges on ionic species to 0.8e[42] to produce results comparable with the experimental density and diffusion coefficient of ions in the neat [BMIM][$PF_6$] electrolytes. In the case of pectin, the LJ nonbonded parameters are directly taken from the GLYCAM06J parameter set.[43] Quantum calculations were performed to generate intramolecular bonded interaction parameters (except the dihedrals) and partial charges for all atoms of pectin, as discussed in the following **Section**. The Lorentz-Berthelot arithmetic rules ($\sigma_{ij} = (\sigma_i + \sigma_j)/2$ and $\epsilon_{ij} = (\epsilon_i\epsilon_j)^{1/2}$) were followed to calculate the nonbonded cross-interaction terms between different atom types of pectin. However, geometric mixing rules ($\sigma_{ij} = (\sigma_i\sigma_j)^{1/2}$ and $\epsilon_{ij} = (\epsilon_i\epsilon_j)^{1/2}$) were used for cross-terms in BMIM-$PF_6$. To be consistent with the OPLS-AA force field, the cross terms between atoms of pectin and BMIM-$PF_6$ were also calculated using the geometric mixing rules. The molecular structure of BMIM, $PF_6$, and pectin are shown in **Figure 1.**

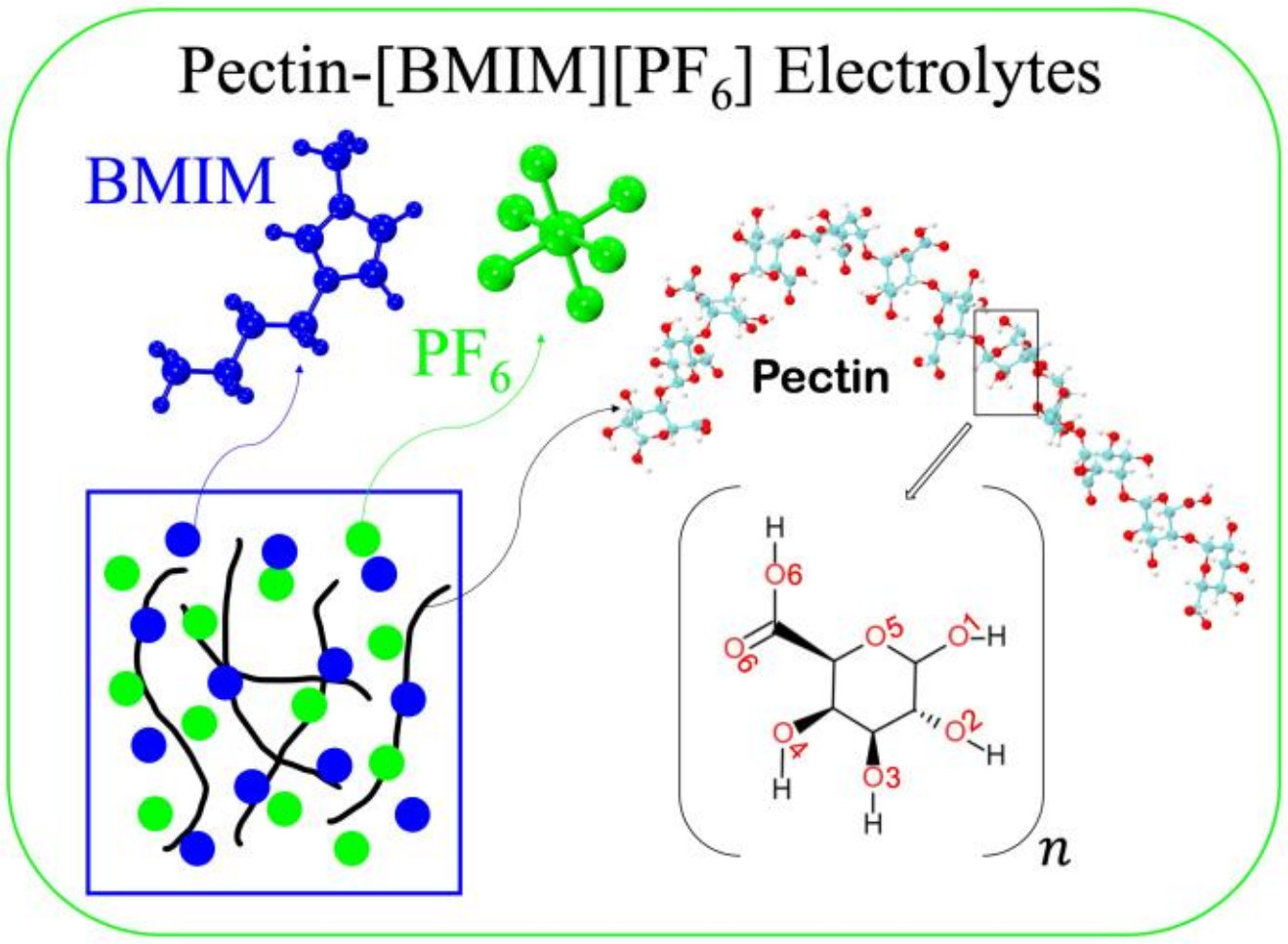

**Figure 1**. Schematic representation of the pectin-IL electrolyte systems simulated in this work and the chemical structure of the molecules BMIM, and $PF_6$, pectin polymer, and the 2D structural representation of monomer of pectin.

## Development of intramolecular force field parameters for pectin

To generate the intramolecular force field parameters for pectin, quantum mechanical calculations were performed by using the Gaussian16 package.[44] We simulated a single monomer of pectin using density functional theory (DFT) with the B3LYP/6-311g** basis set.[45,46] To obtain the equilibrium bond lengths and angles, we optimized the geometry of a monomeric unit of pectin. The force constants of the harmonic potentials for all the bonds and angles were estimated by the normal mode analysis of the vibrational frequencies.[47] The partial charge of all atom types of pectin was evaluated by using the restrained electrostatic potential (RESP) fitting method.[48] The parameters of equilibrium dihedral torsion angles and the respective force constants are directly taken from the GLYCAM06J parameter set.[43] The intramolecular force field parameters developed in this paper along with the partial atomic charges of pectin are provided in the **Table ST1** in the supplementary information (SI).

## The system setup and equilibration

The initial structure of the pectin polymer was built by linking the D-galacturonic acid with 1-4 alpha linkage using the GLYCAM package.[43] The prepared polymer was then esterified with the COOH group and terminated with the hydroxyl group. An appropriate number of pectin polymer chains was added randomly to pure [BMIM][$PF_6$] in a simulation box by using PACKMOL[49] to build pectin-[BMIM][$PF_6$] electrolytes at different weight percentages of pectin. Since the molecules are randomly added in the simulation box with rotational and translational degrees of freedom, the initial density is much lower than the experimental densities. A schematic representation of the simulated pectin-IL electrolyte systems and the chemical structure of the molecules are presented in **Figure 1**. The details of the systems at different loading of pectin chains are provided in **Table ST2** in the (SI). The conformation of the entire system, consisting of randomly inserted pectin and ionic liquid molecules, was minimized by employing the steepest-descent method with a cutoff force of $10^3$ kJ/mol/nm.[50] The minimized structures were then subjected to an NVT ensemble using the V-rescale thermostat with a coupling constant of 1 ps.[51] After that, we performed an NPT equilibration using the Berendsen thermostat[52] with a temperature coupling constant of 1 ps and the Parrinello-Rahman barostat with a pressure coupling constant of 2 ps, respectively.[53] A compressibility factor of $4.5 \times 10^{-5}$ $bar^{-1}$ was used for NPT equilibration during which the density was equilibrated. Production runs were conducted with the time step of 1 fs in an NPT ensemble to generate MD trajectories of 300 ns at an elevated temperature of 425 K. Periodic boundary conditions (PBC) were applied in all three directions. The integration of the equations of motion was carried out by using the leap-frog algorithm[54] and the trajectories were saved every 1 ps. The bond lengths involving hydrogen atoms are constrained with the LINCS algorithm.[55]

The simulations very clearly show that pectin is soluble in [BMIM][$PF_6$], not only at low loading but also at high loadings (see **Figure S1**). Conversely, pectin possesses excellent ion solvation capabilities similar to polyethylene oxide, a widely studied polymer for battery applications. The snapshots taken from equilibrated configurations of the pectin-IL electrolytes at 2 and 20 wt. %s of pectin loading shows excellent uniform distribution of ions throughout the simulation box. We note that in molecular dynamics simulations of systems containing polymers, different initial configurations can lead to different results if the equilibrium is not achieved correctly. For such systems, it is desirable to bring the system to equilibrium with advanced techniques to avoid potential energy traps, for example, by employing annealing or replica exchange techniques. Our simulations verify the dependency on initial configurations and equilibration protocol, as discussed in **Section S2** of the SI. In sum, using the protocol employed in this paper, the obtained results and conclusions are found to be independent of the initial configuration and a different equilibration protocol. A detailed discussion on the effect of equilibration protocol and dependency on initial configurations is provided in the SI (see **Figures S2** and **S3**).

# RESULTS AND DISCUSSION

**Diffusion coefficients of cations and anions**

To understand the ion transport properties, we calculated the mean squared displacements of BMIM and $PF_6$ of the pectin-IL electrolytes at different loadings of pectin. The mean squared displacement is calculated using the formula, $\left\langle \left(\boldsymbol{R}(t+t') - \boldsymbol{R}(t')\right)^2 \right\rangle_{t'}$, where $\boldsymbol{R}$ is the position vector of ions and $\langle \cdots \rangle$ represents the ensemble average over the number of ions and all possible time origins, $t'$.

The results displayed in **Figure 2** demonstrate a consistent decrease in the MSDs as pectin loading increases, with more pronounced effects observed at higher weight percentages. The curves of the MSD display subdiffusive behavior for the ions up to a few nanoseconds, followed by an apparent linear behavior with time (i.e., apparent diffusive regime) after 10s of ns. To check the linearity, we fitted the MSD curves to a power law, $\left\langle \left(\boldsymbol{R}(t+t') - \boldsymbol{R}(t')\right)^2 \right\rangle_{t'} \sim t^{\beta}$ beyond the apparent diffusive regime, and the corresponding time dependent exponents are provided in **Figure S4**. These exponents suggest that our simulation trajectories were sufficiently long to capture the diffusive behavior of the ionic species. The MSD results are affected uniformly for both the ionic species, and the decrease is linear with the loading at higher wt. %s (see inset to **Figure 2**).

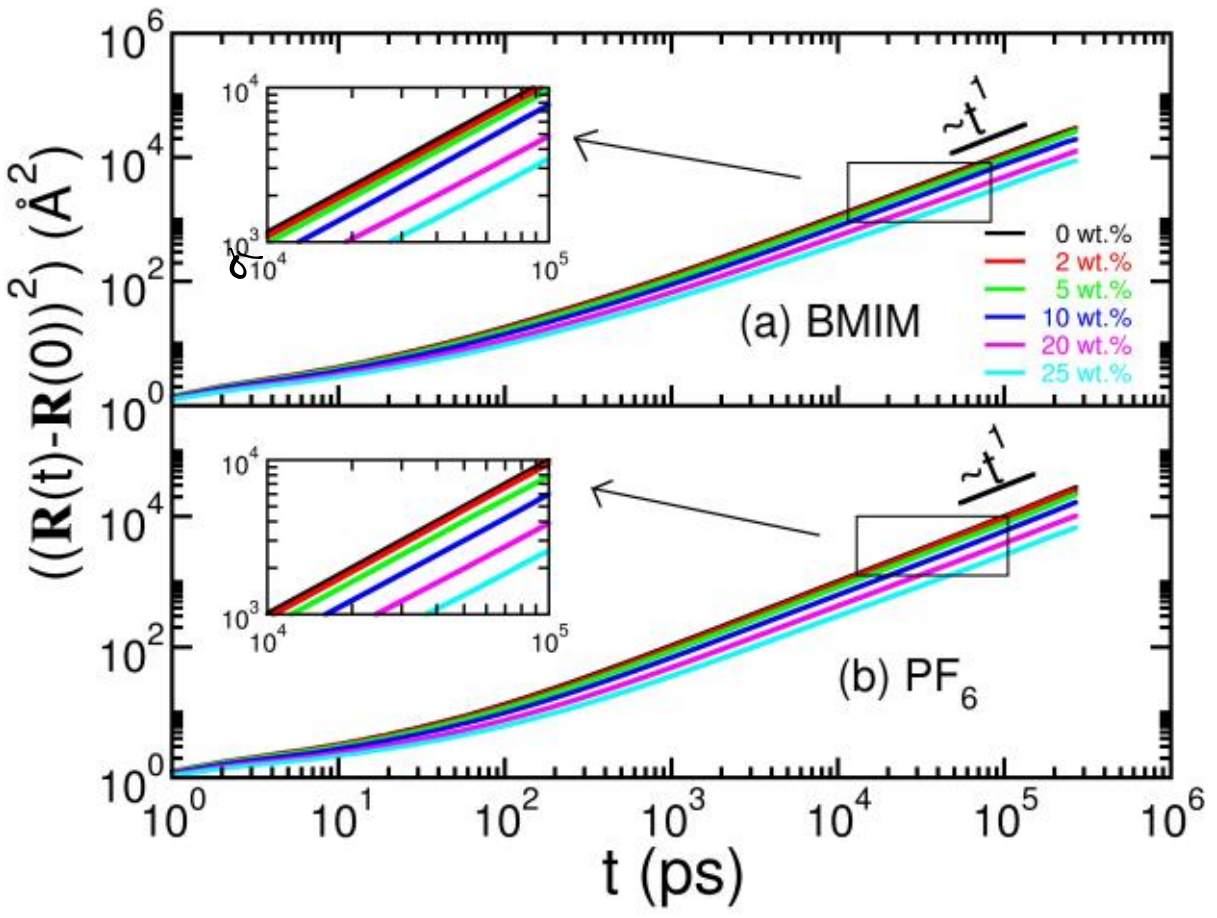

**Figure 2**: Mean squared displacements of (a) cations and (b) anions at different wt. %s of pectin in pectin-IL electrolytes.

We used Einstein's relation to calculate the self-diffusion coefficient of the ions in the diffusive regions, $D = \lim_{t\to\infty} \left\langle \left(\boldsymbol{R}(t) - \boldsymbol{R}(0)\right)^2 \right\rangle / 6t$ and the results are presented in **Figure 3**. To ascertain the diffusive regime of MSDs in our simulations, we fitted the MSD data to a power-law relation, MSD$\sim t^{\beta}$ and the respective exponent values, $\beta$ are provided in the **Table 1**.

| wt. % | $\boldsymbol{\beta}$ **for BMIM** | $\boldsymbol{\beta}$ **for $PF_6$** |
|---|---|---|
| 0 | 1.013 | 0.990 |
| 2 | 1.026 | 1.005 |
| 5 | 1.003 | 1.025 |
| 10 | 1.011 | 1.003 |
| 20 | 0.945 | 1.004 |
| 25 | 0.971 | 0.970 |

**Table 1:** The exponent values $\beta$ of MSD curve obtained by keeping $\beta$ as a free parameter which fitting the MSDs to $\left\langle \left(\boldsymbol{R}(t+t') - \boldsymbol{R}(t')\right)^2 \right\rangle_{t'} \sim t^{\beta}$.

Adding pectin chains in the electrolyte was found to reduce the self-diffusivity of ionic species. It is worth noting that the diffusion coefficient of BMIM is higher than that of $PF_6$ at any loading of pectin because of the electrostatic charge distribution over a larger molecular size. The rate of change of $D$ with the loading is uniform for both the ionic species despite the differences between cation-pectin and anion-pectin interactions. The normalized quantity, $D/D_0$, where $D_0$ is the diffusion coefficient at 0 wt. % loading (i.e., neat IL), shows that the diffusion coefficient drops by only a factor of 4 for the highest loading. This is a significant result because the pectin slightly lowers $D$ while offering multiple benefits to the battery electrolytes, including biocompatibility and improved electrochemical and mechanical stability.

The reduction in diffusion at a loading of 25 wt. % by only a factor of four in pectin-IL electrolytes is a remarkable result when compared to the pectin-EC-LiTFSI electrolytes,[24] where the diffusion coefficient reduces by a factor of a few 100s for a similar loading of pectin [see **Figure S5**]. Further, the pectin-IL electrolytes are beneficial in many ways compared to either neat IL electrolytes or pectin-EC-LiTFSI electrolytes: (i) adding pectin results in disrupted ionic aggregates in IL electrolytes which is very good for efficient ionic conductivity (see next **Section**), (ii) pectin improves the mechanical stability, and (iii) because of its biological nature, pectin makes pectin-IL electrolytes biodegradable. All these benefits come only at the expense of ionic diffusivities by a factor of four for the highest loading (i.e., 25 wt. %) considered in our simulations. The electronegative units in the pectin polymer create a weak associating site for cations, promoting ion hopping and making the pectin chain a viable pathway for ion transport. The chemical structure of pectin is shown the **Figure 1**. Despite the cation's large molecular size, the higher diffusion coefficient of the cation than the counterion suggests that pectin provides an easy pathway for ion movement in IL electrolytes.

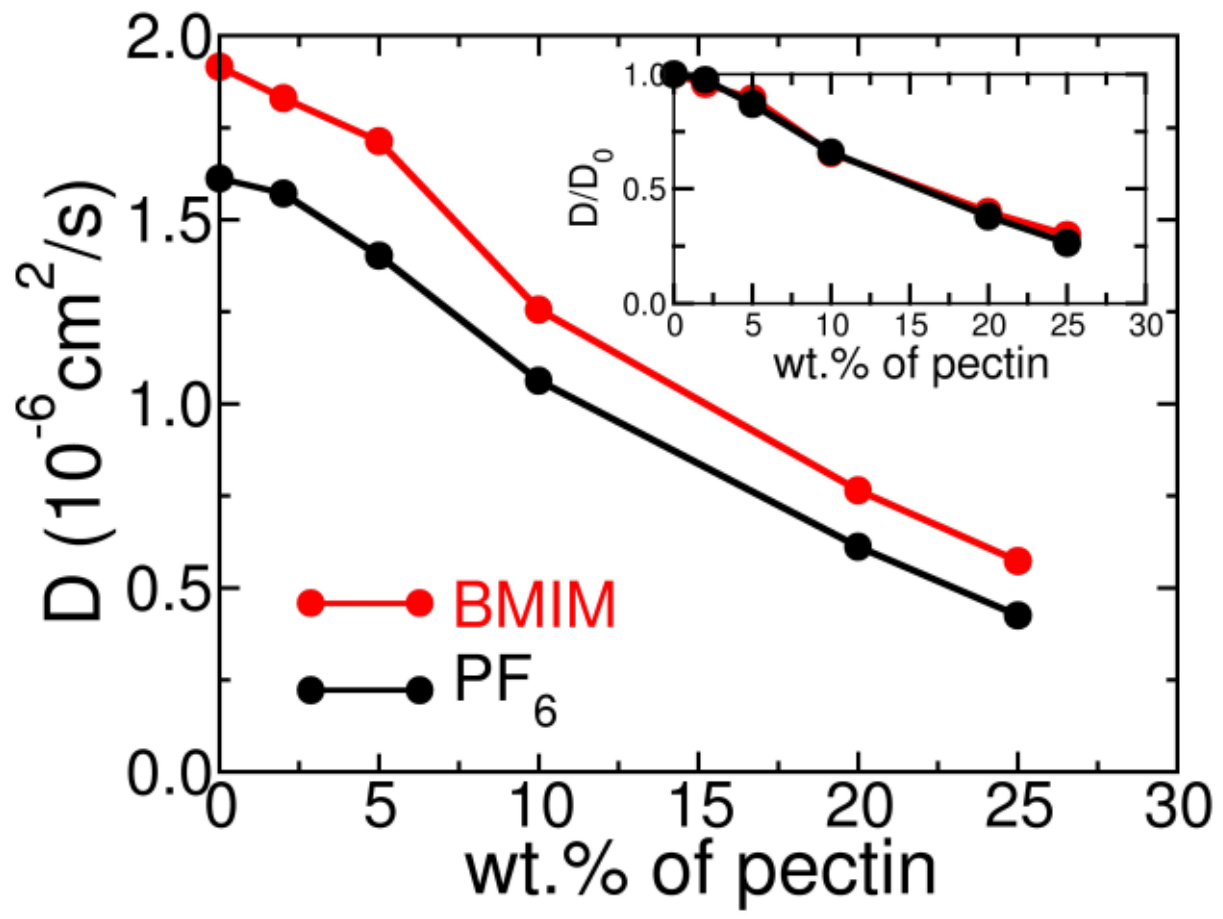

**Figure 3**: Diffusion coefficient of BMIM and $PF_6$ ions as a function of the wt. % of pectin in IL electrolyte. The inset shows the diffusion coefficient of ions scaled with the value of neat IL electrolyte.

The above results present an intriguing advantage of using pectin-IL electrolytes compared to the pectin-EC-LiTFSI electrolytes: introducing pectin into IL electrolytes only moderately affects ionic diffusion. In contrast, the cation diffusion is drastically reduced in the case of pectin-EC-LiTFSI electrolytes.[24] This stark difference can be attributed to the strong electrostatic interactions between the lithium ions and the electronegative groups in the pectin chains. More explicitly, pectin has many negatively charged carboxyl and carboxylate groups. These groups can form strong electrostatic interactions with the positively charged lithium ions, forming strong lithium-polymer association pairs, thus decreasing the lithium diffusion more dramatically. The reduction in the diffusion coefficient implies an increase in viscosity, assuming a direct correlation between D and $\eta$, but further investigation is required to establish such a mechanism.

The structural changes of a battery electrolyte play a crucial role in its efficiency, and pectin can affect these changes. In the following **Section**, we examine how pectin specifically impacts the electrolyte structure.

**Radial distribution functions, coordination numbers, and ion association probabilities**

We computed the radial distribution function (g(r)) and coordination numbers (CN(r)) between different atomic pairs, including cation-anion, cation-polymer, and anion-polymer, to understand the structural changes in the electrolyte and the implications to ion-polymer interactions. Specifically, we examined the N(BMIM)–P($PF_6$), N(BMIM)–O(pectin) (the oxygens of pectin polymers are shown in **Figure 1** in the SI), and P($PF_6$)–O(pectin) pairs, among others. These pairs were chosen because they are important in understanding the structural changes and ion-polymer interactions in the electrolytes. The g(r) of cation-anion pairs showed a strong peak at 4.75 Å, followed by the first minimum at 7.5 Å (see **Figure 4(a)**). The local ordering of cation and anion structure disappears beyond length scales of 12 Å, settling the isotropic structural organization. The first maximum and minimum position is unaffected by pectin, and the overall shape of the g(r) remains unaffected. However, the first maximum value slightly increases with pectin loading. Despite this slight change, we notice the coordination numbers to decrease monotonically with the loading at any distance $r$. The CN(r) within the first coordination shell plotted in the inset of **Figure 4(b)** displays a linearly decreasing trend. This result is promising since it directly implies an increase in either free ions or smaller ionic aggregates with the loading of pectin biopolymer.

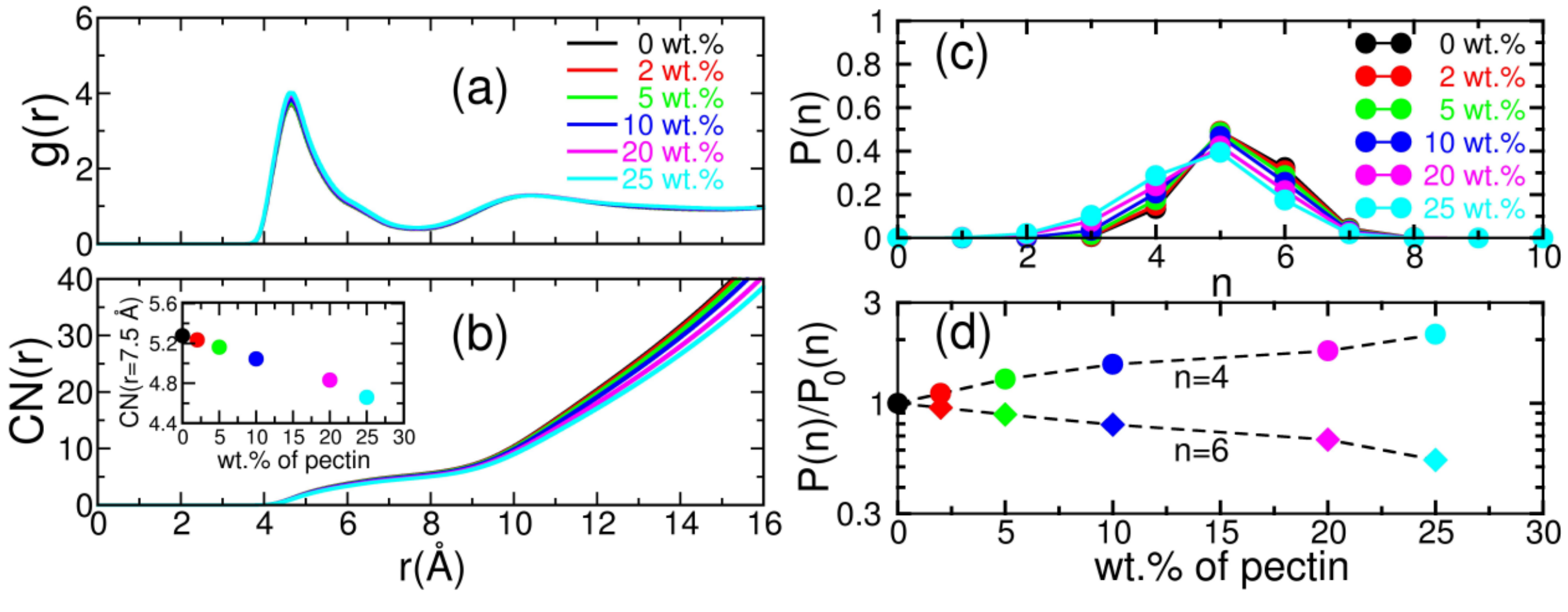

**Figure 4**: (a) Radial distribution function of cation-anion pairs, (b) Coordination number of anions around cations, (c) The probability of finding $n$ number of $PF_6$ ions within the first coordination shell of BMIM

ions, and (d) Relative change in P(n) for n = 4 and n = 6, highlighting the effect of pectin on the ion association probabilities.

To establish the role of pectin on the formation and promotion of smaller ionic aggregates, we calculated the probability of ion association function, P(n), between various atomic pairs. The ion association probability is calculated using,

$$\mathrm{P(n)} = \frac{1}{N_{\mathrm{frames}}} \sum_{i=1}^{N_{\mathrm{frames}}} \sum_{j=1}^{N_{\mathrm{ions}}} \frac{\delta_{n_i n_j}}{N_{\mathrm{ions}}}, ------ (2)$$

where $\delta_{n_i n_j}$ is the Kronecker delta function to count $n$ number of counterions within the first coordination shell of the ion $j$, $N_{\mathrm{ions}}$ is the total number of ions, $N_{\mathrm{frames}}$ is the total number of frames. Here, $\delta_{n_i n_j}$ is defined in such a way that $\delta_{n_i n_j} = 1$ if $n_i^{\mathrm{th}}$ ion is found within the first coordination shell of $n_j^{\mathrm{th}}$ ion, and $\delta_{n_i n_j} = 0$, otherwise. The quantity, P(n) calculated using the above equation accounts for the ionic aggregates containing 1 cation(anion) and n neighbor anions(cations) in the first coordination shell. While this definition does not entirely capture all possible aggregates (for example, aggregates containing more than 1 cation and more than 1 anion simultaneously in an aggregate), P(n) captures the most important statistics related to small aggregates and free ions, which are crucial for ionic conductivity. However, we focused on the P(n) for $PF_6$ ions around BMIM ions, as shown in **Figure 4(c)**.

The analysis of P(n) primarily confirms two main observations. Firstly, the peak in the P(n) distribution is observed at $n = 5$; i.e., there are approximately 5 anions within the first coordination shell of cations, consistent with previous literature on neat IL electrolytes.[38] However, we found that the ion association probability P(n = 5) decreases with increasing loading of pectin. This suggests that pectin disrupts the formation of larger ionic aggregates around cations.

Secondly, we compared the P(n = 4) and P(n = 6) distributions to investigate the formation of smaller ionic aggregates. Our results indicate that the formation of larger ionic aggregates is more prominent in neat and low pectin-containing pectin-IL electrolytes. However, at high loadings of pectin, the smaller ionic aggregates are dominant (see **Figure 4(d)**). The decrease in the value of P(5) is significant, which demonstrates a reduction of 20% with the addition of pectin at the highest wt. % (see **Table 2**). This interprets that the addition of pectin provides more accessible pathways for ions. However, P(6) decreases by 45%, which means already half of the total counterions decreases with the addition of pectin. These findings reveal that the higher ionic clusters are broken into smaller ionic clusters. Since the results point out that the smaller ionic aggregates are promoted over the large-sized ones, we may expect quantities like P(1), P(2) to increase more efficiently at high loadings of pectin for wt. %s higher than 25. For such scenarios, P(4) is likely to show a plateau with the loading of pectin and may even decrease at very large values of pectin loading. The reduction in the probability of counterions around the cation is attributed to the dilution effect due to pectin. Moreover, the larger ionic clusters are decreasing to smaller ionic clusters, which indicate that the pectin can give more pathways for ions. This implies that the presence of pectin promotes the formation of free ions or smaller ionic aggregates in pectin-IL electrolytes.

| **wt. %** | $\boldsymbol{P(4)/P_0(4)}$ | $\boldsymbol{P(5)/P_0(5)}$ | $\boldsymbol{P(6)/P_0(6)}$ |
|---|---|---|---|
| 0 | 1.00 | 1.00 | 1.00 |
| 2 | 1.11 | 1.00 | 0.95 |
| 5 | 1.30 | 0.99 | 0.88 |
| 10 | 1.53 | 0.95 | 0.79 |
| 20 | 1.77 | 0.86 | 0.67 |
| 25 | 2.12 | 0.80 | 0.54 |

**Table 2**: Ion association probabilities of finding $n$ number of anions around cations, $P(n)$ at $n = 4$, 5 and 6, normalized with respect to the pectin-free neat electrolyte.

It is widely recognized that certain types of SPEs exhibit low ionic conductivity at lower salt concentrations, which then increases to its maximum at a specific optimal concentration.[56] However, once the maximum conductivity is reached, the conductivity begins to decrease at higher salt concentrations due to the formation of large-sized clusters. The pathways for the ions are blocked by the large-sized clusters.[57–59] In general, smaller ionic aggregates lead to higher diffusion and higher ionic conductivity in ionic liquids[60,61] because smaller aggregates have a lower degree of structural hindrance and less restricted movement compared to larger aggregates. Therefore, the presence of smaller ion aggregates plays a significant role in improving the overall ionic conductivity. The results of the $\mathrm{P(n)}$ analysis are consistent with the coordination number of anions around cations. Overall, our findings suggest that adding pectin chains to the electrolyte system provides more accessible pathways for ions, which can increase the ionic conductivity of the electrolytes.

So far, we have presented the analysis based on the results of g(r) and P(n) cation-anion pairs, and it is also important to understand nature of ion interaction with the polymer. To understand the interactions between the ions and polymer at a qualitative level, we calculated the and CN(r) between the ions and the oxygens of the pectin. First of all, the results of **Figure 5** reveal that the ion-polymer interactions are much weaker than that of cation-anion interactions as seen through the respective peaks in g(r). This result indicates that cation-anion interactions are more important in pectin-IL electrolytes and that they could also eventually influence the motion of ionic species and the ion transport. Further, the differences between the cation-polymer and anion-polymer interactions are indicated by the extent of the first coordination. Specifically, the first peak occurs at 4.2 Å for cation-polymer pairs compared to 5.6 Å for anion-polymer pairs. Similarly, the first minima occur at 5.0 Å for cation-polymer pairs compared to 7.3 Å for anion-polymer pairs which define the extent of the first coordination shell. These results establish that cations exhibit more preferential interaction with polymer than anions. Because of the increased coordination numbers around the polymer with the loading (see **Figure 5**(c-d)), the cation-anion interactions would decrease (or vice-versa) (see **Figure 4**(b)) which explains the higher tendency of smaller ionic aggregate formation over the larger ones with the loading of pectin in IL electrolytes. The corresponding results of P(n) also presented in **Figure S6**.

Based on these results, we propose that the diffusion of ionic species is more likely to be influence by the cation-anion correlations than the effect coming directly from the polymer. Further, we also propose that the ion-pair correlations to be more closely coupled to the anion diffusion because of the large extent of the first coordination shell. We examine these mechanisms in detail in the following two **Sections**.

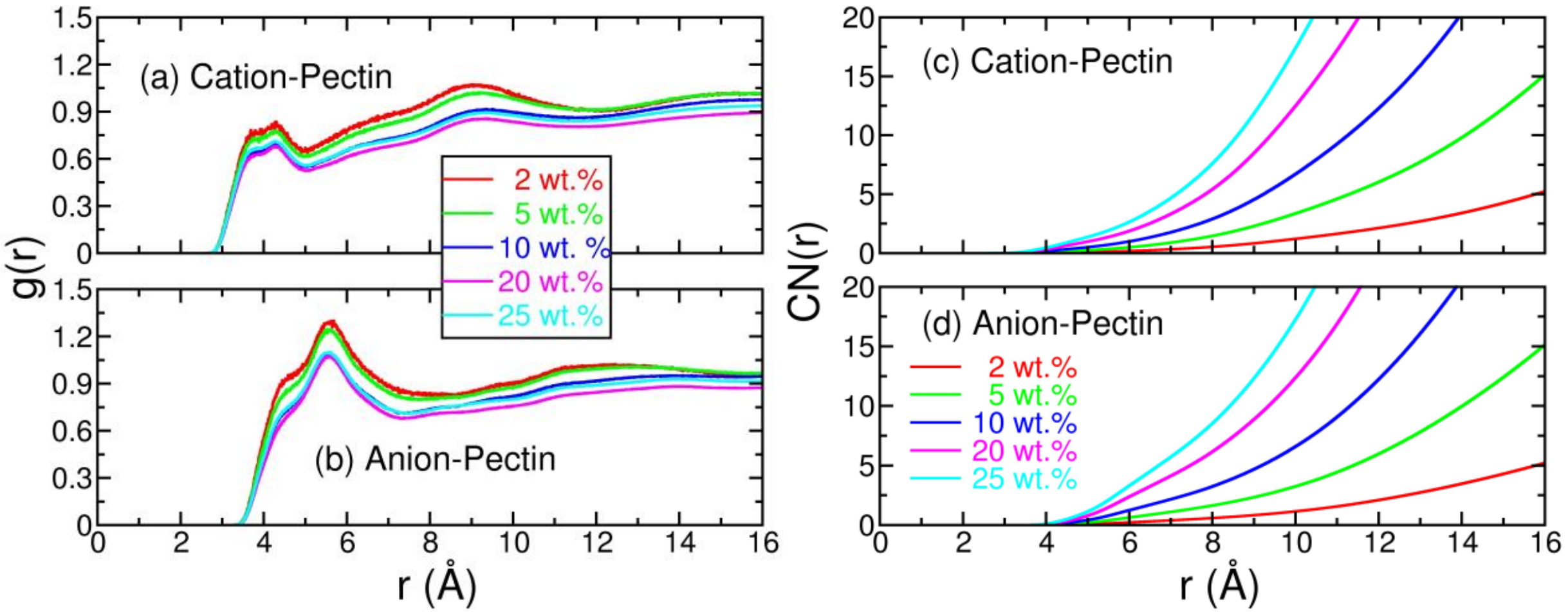

**Figure 5**: The radial distribution functions representing the (a) cation-polymer pairs and (b) anion-polymer pairs and the (c-d) respective coordination numbers. The O atoms were chosen as the representatives of the pectin chain due to the electronegativity of oxygen atoms in pectin.

Our analysis demonstrates that pectin induces structural changes in the electrolyte that disrupt large ionic clusters and promote the formation of smaller ionic clusters. These structural changes may potentially increase the ionic conductivity and the mechanical stability of the pectin-IL electrolyte, particularly with increasing loading of pectin polymer. Therefore, the observed changes in the electrolyte's structural properties can directly affect the relaxation phenomena of ion-pairs in the pectin-IL electrolyte. To further investigate the impact of pectin on ion-pair relaxation, we analyzed the ion-pair correlation functions for various pectin loadings. The results of our analysis are presented in the next **Section.**

**Ion-pair autocorrelation functions**

To investigate the relaxation behavior of ion-pairs in the pectin-IL electrolyte, we calculated the ion-pair correlation function, $C(t)$, from 300 ns long trajectories. The ion-pair correlation function $C(t)$ is calculated as $C(t) = \langle h(t)h(0)\rangle / \langle h(0)h(0)\rangle$, where the angular bracket $\langle \cdots \rangle$ represents the ensemble average over all the ion-pairs and all possible time origins. In our calculation, the population variable $h(t)$ takes a value of 1 if an ion-pair is intact, meaning that the cation and anion are within a cutoff distance of 7.5 Å where the radial distribution function, $g(r)$, shows the first minima. Otherwise, $h(t)$ takes a value of zero. The ion-pair correlation function provides insight into the relaxation behavior of ion-pairs in the electrolyte. Specifically, it measures the probability of an ion-pair being intact at a given time, provided it is intact at the initial time, $t = 0$.

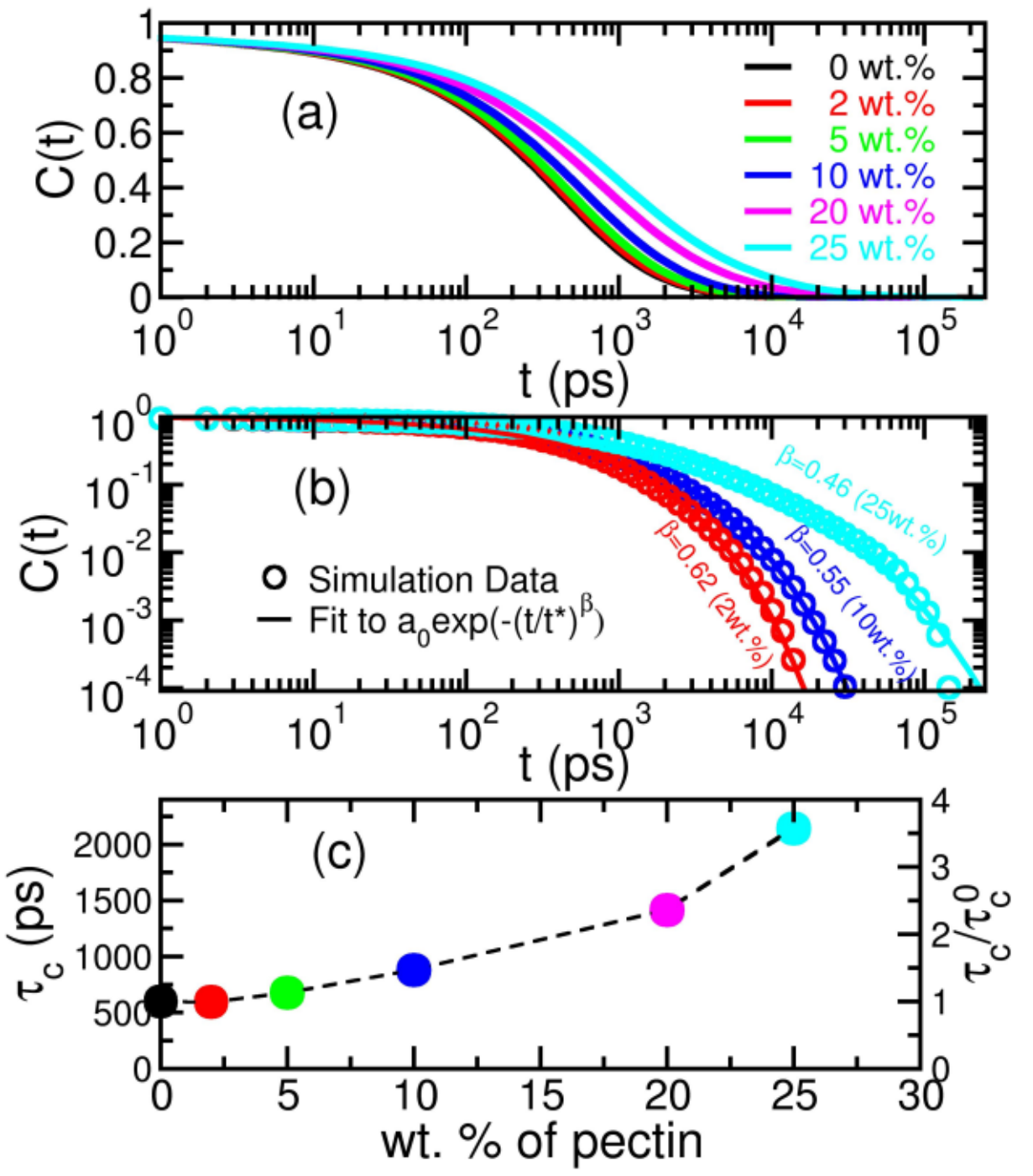

**Figure 6**: (a) The ion-pair correlation function $C(t)$ at different loadings of the pectin, (b) log-log plot of $C(t)$ with fits accordingly to $a_0 \exp\left(-(t/t^*)^{\beta}\right)$ at selected wt. %s of pectin, i.e., 2, 10 and 25, and (c) the mean relaxation times calculated using the formula $\tau_c = \int_0^{\infty} a_0 \exp\left(-(t/t^*)^{\beta}\right) \mathrm{dt} = a_0 t^* \Gamma(1 + 1/\beta)$ indicated on the left vertical axis, and $\tau_c$ normalized by its values in neat IL electrolytes is shown on the right vertical axis.

The results presented in **Figure 6** reveal a slow decay of $C(t)$ during the initial timescales of up to 100 ps. During these timescales, the cations and anions are strongly associated with each other, with the structural reorganization phenomenon being the most dominant phenomenon occurring in the first coordination shell. As the structural relaxation (i.e., ions diffusing beyond the extent of the first coordination shell) becomes dominant slowly after the initial timescales, the ion-pairs dissociate rapidly. The rate of decay of $C(t)$ slows down monotonically with an increase in pectin loading, and complete dissociation of ion-pairs is observed on timescales of the order of $10^3$ to $10^4$ ps depending on the pectin loading. We fitted the curves with a stretched exponential function, the respective simulation data for C(t), and the fitted curves at selected wt. %s of pectin, i.e., 2, 10 and 25, are shown in **Figure 6(b)**. The observed $\beta$ values range from 0.45 for the highest loading of pectin to 0.65 for the lowest pectin loading, indicating the degree of stretching of the exponential decay for the highest loading. The fitting quality is quite good, with the correlation coefficient of the fits observed to be in the range of 0.95 to 0.99.

We plotted the relaxation timescales in **Figure 6(c)**, which show a consistent increase in the ion-pair relaxation timescale with the addition of pectin into IL electrolytes. For the highest loading, the ion-pair relaxation times increase by a factor of 4 compared to the neat IL electrolytes (see ~~inset to~~ **Figure 6(c)**). The increase in the relaxation times implies a slow movement of ions in the electrolyte matrix. Therefore, as the structural relaxation dynamics are slowed down qualitatively, we expect a long time for the ions to diffuse beyond the extent of the first coordination shell.

**Correlation between diffusion coefficients and ion-pair relaxation timescale**

The results presented in the previous **Section** suggest a direct relationship between the diffusion coefficients and the ion-pair relaxation timescales. To better understand the correlations, we plot $D$ of cations and anions against $\tau_c$ in **Figure 7**. Consistent with the above conjecture, we find that the diffusion decreases with an increase in $\tau_c$ in a power-law-like behavior. The validity of diffusivity power-law relation $D \sim \tau_c^{-1}$ explains the ability to design an electrolyte to increase the ionic conductivity of the electrolyte at the expense of its mechanical strength. This power-law relation is commonly valid in liquid and IL electrolytes, in which the relevant timescale can be simply tuned to influence the diffusivity to attain high ionic conductivity. The results were fitted to $D \sim \tau_c^{-\lambda}$ where $\lambda$ quantifies the degree of correlation between $D$ and $\tau_c$.

The obtained exponents are 0.75 for cations and 0.82 for anions with the respective power-law equations, $D \sim \tau_c^{-0.75}$ for cations and $D \sim \tau_c^{-0.82}$ for anions. These results support our proposals (based on the results of **Figure 4** and **Figure 5**) that (i) the diffusion of ionic species is more likely to be influence by the cation-anion correlations than the effect coming directly from the polymer and (ii) the ion-pair correlations to be more closely coupled to the anion diffusion because of the large extent of the first coordination shell. However, the above results are quite distinct compared to the transport mechanisms reported for pectin-EC-LiTFSI electrolytes[24] and polymeric ionic liquid electrolytes.[38,62] Unlike the cations that interact with both anions and polymer, the anions interact primarily with the cations in the electrolyte. Therefore, the degree of correlation of anions is closer to a one-to-one correspondence-like relation than the degree of correlation obtained for cations, consistent with the results of g(r) for ion-polymer interactions (**Figure 5**(a-b)). Specifically, since the anions have less preferential interaction with the polymer compared to the cations, their diffusion mechanisms to be less distracted by the effects coming from polymer.

The above results show that the structural relaxations occurring via the formation and dissociation of ion-pair are the primary ion transport mechanism in pectin-IL electrolytes. However, the exponents being less than 1 also indicate a possible secondary mechanism of ion transport, the analysis of which is beyond the scope of this paper. We speculate that an approximately 20% of the contribution to the overall ion transport mechanism comes from a different one other than the ion-pair structural relaxation (i.e., the formation and dissociation mechanism of ion-pairs). According to the power law relation, i.e., $D \sim \tau_c^{-1}$, explains the dependency of diffusivity on the ion-pair relaxation so that the $\tau_c$ can be simply tuned to influence the diffusivity to achieve an electrolyte with higher ionic conductivity. We can design an electrolyte on the concept of $D \sim \tau_c^{-1}$ to achieve an electrolyte with higher ionic conductivity and viscosity. We would expect that pectin can be responsible for the increase in viscosity due to its highly viscous nature. Again the $P(5)$ and $P(6)$ result revealed that the smaller ion aggregates are encouraged with the pectin loadings. Hence, pectin is a potential candidate to increase ionic conductivity and viscosity. Further research is needed to understand the complete mechanisms behind the behavior of pectin-[BMIM][$PF_6$] electrolytes to optimize their properties for use in energy storage devices.

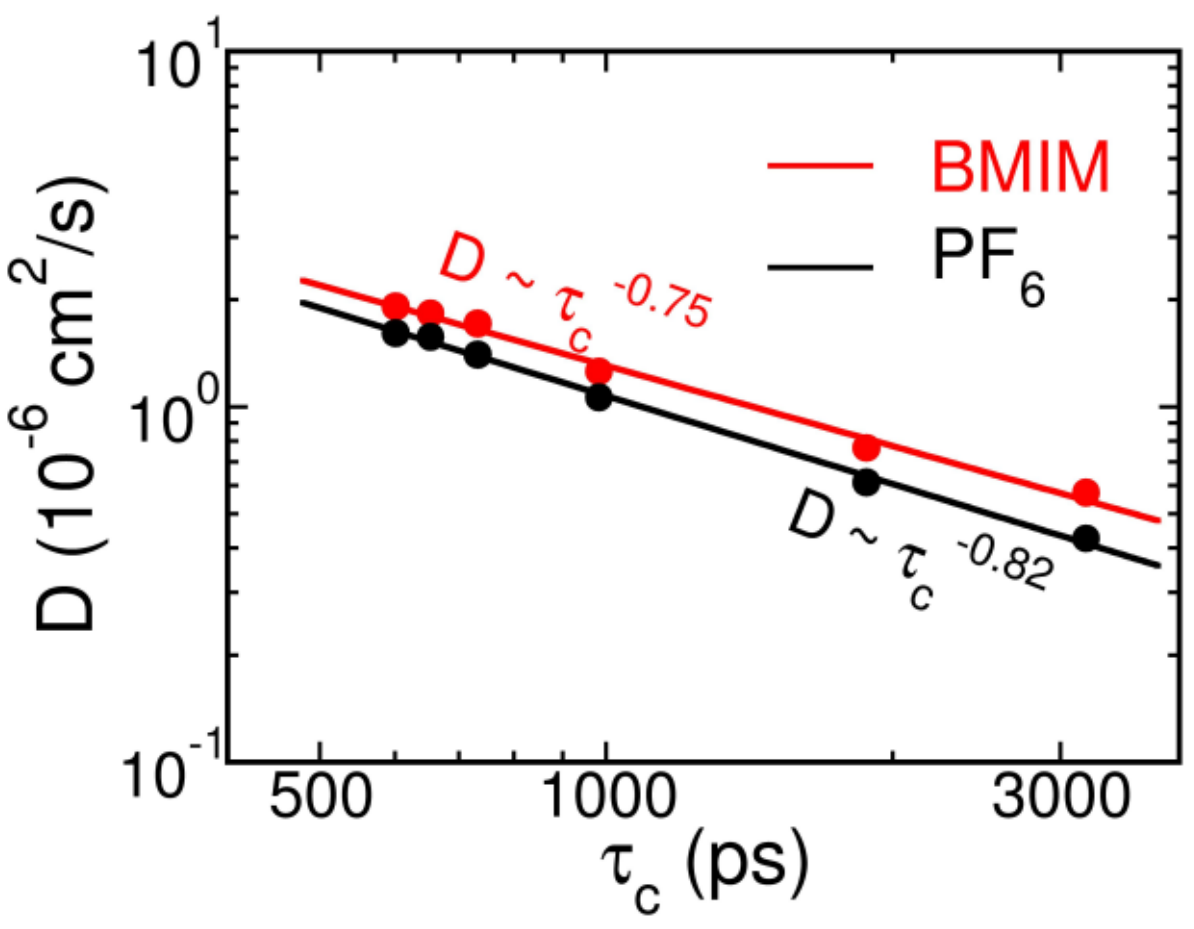

**Figure 7**: Correlations between the diffusion coefficient and the ion-pair relaxation timescales.

## CONCLUSIONS

The paper presents the results of molecular dynamics simulations investigating the effect of pectin, a natural biopolymer, on the ion transport mechanisms and structural properties of ionic liquid electrolytes. The simulations are performed for different loadings of pectin in ILs, ranging from 0 to 25 wt. %. The diffusion coefficient of ions, cation-anion radial distribution functions, ion association statistics, and ion-pair relaxation timescales are calculated to investigate the effects of pectin on the transport and structural properties of the pectin-IL electrolytes.

The results indicate that the diffusion coefficients of ions decrease with pectin monotonically, indicating higher resistance to ion transport. Due to the ion-polymer interactions, the coordinating ions around the polymer increase with the loading of pectin, which leads to the reduced mobility of ions. We also investigated the ion-pair time correlation function and found a strong correlation between ion diffusion and relaxation timescales ($D \sim \tau_c^{-0.75}$ for cations and $D \sim \tau_c^{-0.82}$ for anions). Interestingly, the coordination numbers of ions around each other ionic types (i.e., cations around anions and vice-versa) decrease with increasing pectin loadings. Further, our analysis revealed that pectin induces intriguing structural changes so that the availability of smaller ionic aggregates is more than the larger ionic aggregates (i.e., P(n=4)>P(n=6)). This result contradicts the literature on neat ionic liquid electrolytes where the larger ionic aggregates with n=6 are more pronounced, and the ion-ion correlations are dominant. Our simulations and analysis of ion-ion association statistics point to an interesting consequence of pectin-IL electrolytes with huge potential for efficient ionic conductivity induced by pectin.[63–66]

This work highlights an interesting advantage of using pectin-IL electrolytes for rechargeable battery applications compared to the pectin-EC-LiTFSI electrolytes: introducing pectin into IL electrolytes only moderately affects ionic diffusion. Despite the slightly diminished ion diffusivity, the overall benefits of mechanical properties and biodegradability make it a favorable option for practical applications. Based on the results, we speculate that biopolymers can replace synthetic polymers to make the rechargeable battery electrolytes more biocompatible, highly stable, and efficient. The findings of this study provide insights into the potential use of pectin as a sustainable and biodegradable alternative to traditional solid polymer electrolytes for energy storage systems. Our results suggest that pectin-based electrolytes can provide a diffusive channel for ions, leading to improved ion transport and mechanical stability.

## SUPPLEMENTARY MATERIAL

The supplementary information contains: **(S1)** The potential model and interaction potential, **(Table ST1)** The force field parameters for pectin polymer, **(Table ST2)** System details, **(S2)** The number density profile of cations and anions in x, y, and z directions, **(S3)** Effect of equilibration protocols and dependency on initial configurations, **(S4)** The exponent β in MSD of BMIM and $PF_6$ ion, **(Table ST3)** The exponent values of the MSD in diffusive regions, **(S5)** The normalized ionic diffusivity comparison between pectin-IL and pectin-EC-LiTFSI, and (**S6**) Association probability of different pair of ions.

## ACKNOWLEDGEMENTS

The authors acknowledge the Computer Center of IIT Jodhpur, the HPC center at the Department of Physics, Freie Universität Berlin (10.17169/refubium-26754), for providing computing resources that have contributed to the research results reported in this paper. RRN acknowledges the support by Deutsche Forschungsgemeinschaft, Grant No. CRC 1349, Code No. 387284271, Project No. C04. SM acknowledges support for the SERB International Research Experience Fellowship SIR/2022/000786 and SERB CRG/2019/000106 provided by the Science and Engineering Research Board, Department of Science and Technology, India.

## AUTHOR DECLARATIONS

### Conflict of Interest

The authors have no conflicts to disclose.

### Author Contributions

**Sipra Mohapatra:** Simulations, Methodology, Data analysis and writing first draft and revisions;
**Sougata Halder:** Simulations, Data analysis and contributing to writing first draft;
**Sachin R. Chaudhary:** Data analysis, Review & editing;
**Roland R. Netz:** Review & Editing; Resources; Supervision;
**Santosh Mogurampelly:** Conceptualization; Supervision; Methodology; Resources, Writing review & editing.

## DATA AVAILABILITY

The data that support the findings of this study are available within the article and its supplementary material.